\newcommand{\Tr}{\mathop{\mathrm{Tr}}\nolimits}

\documentclass[twocolumn,eqsecnum,showpacs]{revtex4}

\usepackage{amsfonts,amssymb,amsbsy,bm,graphicx}

\begin{document}

\title{Vectorlike representation of one-dimensional scattering}

\author{Luis L. S\'anchez-Soto}
\affiliation{Department of Physics,
Lakehead University, Thunder Bay,
Ontario P7B 5E1, Canada}

\author{Jos\'e F. Cari\~{n}ena}
\affiliation{Departamento de F\'{\i}sica Te\'orica,
Facultad de Ciencias, Universidad de Zaragoza,
50009 Zaragoza, Spain}

\author{Alberto G. Barriuso and Juan J. Monz\'on}
\affiliation{Departamento de \'Optica,
Facultad de F\'{\i}sica,
Universidad Complutense,
28040 Madrid, Spain}

\date{\today}

\begin{abstract}
We present a self-contained discussion of
the use of the transfer-matrix formalism
to study one-dimensional scattering. We
elaborate on the geometrical interpretation
of this transfer matrix as a conformal mapping
on the unit disk. By generalizing to the unit
disk the idea of turns, introduced by Hamilton
to represent rotations on the sphere, we
develop a method to represent transfer
matrices by hyperbolic turns, which can be
composed by a simple parallelogramlike rule.
\end{abstract}


\maketitle

\section{Introduction}

The quantum mechanics of one-dimensional
scattering describes many actual physical
phenomena to a good approximation. In
consequence, this topic continues as an
active line of research with strong
implications both in
fundamental~\cite{Pere83,Bian94,Bian95,Chebo96,Viss99,Miya00}
and in more applied aspects, such as
the study of tunnelling phenomena in
superlattices~\cite{Tsu73,Esak86,Haug89},
to cite only a representative example.

Apart from this interest in research,
scattering in one dimension is also appealing
from a pedagogical point of view and is an
important part in the syllabus of any graduate
course in quantum mechanics. The advantage
of the one-dimensional treatment is that one
does not need special mathematical functions,
while still retaining sufficient complexity to
illustrate physical concepts. It is therefore not
surprising that there have been many didactic
articles dealing with various aspects of such
scattering~\cite{Eber65,Form76,Kama84,Dijk92,Noga96,Barl00,Barl04}.
However, these papers emphasize concepts such
as partial-wave decomposition, Lippmann-Schwinger
integral equations, the transition operator, or
parity-eigenstate representation, paralleling as
much as possible their analogous in two and three
dimensions. In other words, these approaches,
like most if not all the standard textbooks
on the subject~\cite{Gold64,Newt66,Cohe77,Gali90},
employ the formalism of the $S$ matrix.

The elegance and power of the $S$-matrix
formulation is beyond doubt. However, it is
a ``black-box" theory: the system under study
(scatterer) is isolated and is tested through
asymptotic states. This is well suited for
typical experiments in elementary particle
physics, but becomes inadequate as soon
as one couples the system to other.
The most effective technique for
studying such one-dimensional systems
is the transfer matrix, in which the
amplitudes of two fundamental solutions
on either side of a potential cell are
connected by a matrix $\mathbf{M}$.

The transfer matrix is a useful object that
is widely used in the treatment of layered
systems, like superlattices~\cite{Vint91,Webe94,Spru03}
or photonic crystals~\cite{Joan95,Bend96,Tsai98}.
Optics, of course, is a field in which
multilayers are ubiquitous and the
transfer-matrix method is well
established~\cite{Brek60,Lekn87,Yeh88}.
An extensive and up-to-date review of the
applications of the transfer matrix to many
problems in both classical and quantum physics
can be found in Ref.~\cite{Grif01}.

In recent years a number of concepts of geometrical
nature have been introduced to gain further
insights into the behavior of scattering in one
dimension~\cite{Pere01,Yont02,Monz02,Spru04}.
From these analyses it appears advantageous
to view the action of a matrix as a bilinear
(or M\"{o}bius) transformation on the unit
disk. A simple way of characterizing these
transformations is through the study of
the points that they leave invariant.
For example, in Euclidean geometry a
rotation can be characterized by having
only one fixed point, while a translation
has no invariant point. In this paper we
shall reconsider the fixed points of the
bilinear transformation induced by an
arbitrary scatterer, showing that they can
be classified according to the trace of
the transfer matrix has a magnitude lesser
than, greater than, or equal to 2. In fact,
this trace criterion will allow us to classify
the corresponding matrices from a geometrical
perspective as rotations, translations or,
parallel displacements, respectively, which are
the basic isometries (i. e., the transformations
that preserve distance) of the unit disk.

As we have stressed, the advantage of
transfer matrices lies in the fact that they
can be easily composed. Of course, as for any
matrix product, this composition is noncommutative.
A natural question then arises: how this
noncommutativity appears in such a geometrical
scenario? An elegant answer involves the
notion of Hamilton turns~\cite{Hami53,Bied81}.
The turn associated with a rotation of axis
$\hat{\mathbf{n}}$ and angle $\vartheta$ is a
directed arc of length $\vartheta/2$ on the
great circle orthogonal to $\hat{\mathbf{n}}$
on the unit sphere. By means of these objects,
the composition of rotations is described through
a parallelogramlike law: if these turns are
translated on the great circles, until the
head of the arc of the first rotation coincides
with the tail of the arc of the second one,
then the turn between the free tail and the head
is associated with the resultant rotation.
Hamilton turns are thus analogous for spherical
geometry to the sliding vectors in Euclidean
geometry. It is unfortunate that this elegant
idea of Hamilton is not as widely known as it
rightly deserves.

Recently, a generalization of Hamilton
turns to the unit disk has been
developed~\cite{Juar82,Simo89,Barr04}. The
purpose of this paper is precisely to show
how the use of turns affords an intuitive
and visual image of all problems involved
in quantum scattering in one dimension,
and clearly shows the appearance of
hyperbolic geometry in the composition law
of transfer matrices. These geometrical
methods do not offer any inherent advantage
in terms of computational efficiency. Apart
from their beauty, their benefit for the
students lies in the possibility of gaining
insights into the qualitative behavior
of scattering amplitudes, which is important
in developing a physical feeling for this
relevant question.

\section{Superposition principle and transfer matrix}

We consider the quantum scattering in one spatial
dimension by a potential $V(x)$. We assume this
potential to be real (i. e., nonabsorbing) but
otherwise arbitrary in a finite interval $(a, b)$,
and outside this interval, it is taken to be a
constant that we can define to be the zero of
energy. We recall that, because $E>0$, the
spectrum is continuum and we have two linearly
independent solutions for a given value of $E$.
In consequence, the general solution of the
time-independent Sch\"{o}dinger equation for
this problem can be expressed as a superposition
of the right-mover $e^{+i k x}$ and the left-mover
$e^{-ikx}$:
\begin{equation}
\psi(x) =
\left \{
\begin{array}{ll}
A_{+} e^{+ i k (x-a)} + A_{-} e^{- i k (x-a)}
\qquad &  x < a , \\
 & \\
\psi_{ab} (x) & a < x < b \\
 & \\
B_{+} e^{+ i k (x-b)} + B_{-} e^{- i k (x-b)}
& x > b , \\
\end{array}
\right . ,
\end{equation}
where $ k^2 = 2 m E/\hbar^2$ and the subscripts
$+$ and $-$ indicate that the waves propagate
to the right and to the left, respectively
(see Fig.~1). The origins of the movers have
been chosen so as to simplify as much as possible
subsequent calculations.

To complete in a closed form the problem one must
solve the Schr\"odinger equation in $(a, b)$ to
obtain $\psi_{ab}$ and then invoke the appropriate
boundary conditions, involving not only the
continuity of $\psi(x)$, but also of its derivative.
In this way, one obtains two linear relations among
the coefficients $A_\pm$ and $B_\pm$, which can
be solved for any two amplitudes in terms of the
other two, and the result can be expressed as
a matrix equation. The usual choice in most
textbooks is to write the outgoing amplitudes
in terms of the incoming amplitudes (which are
the magnitudes one can externally control)
using the so-called scattering matrix $\mathbf{S}$.
For our purposes in this paper, it will prove
crucial to express a linear relation between
the wave amplitudes on both sides of the scatterer,
namely,
\begin{figure}
\centering
\resizebox{0.80\columnwidth}{!}{\includegraphics{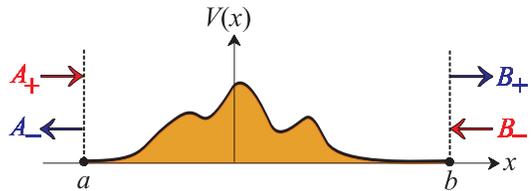}}
\caption{Illustration of the scattering from
an arbitrary potential, showing the input ($A_+$ and
$B_-$) and output ($A_-$ and $B_+$) amplitudes.}
\end{figure}
\begin{equation}
\label{M}
\left (
\begin{array}{c}
A_+ \\
A_-
\end{array}
\right )
=
\mathbf{M}
\left (
\begin{array}{c}
B_+ \\
B_-
\end{array}
\right ) ,
\end{equation}
where $\mathbf{M}$ is the transfer matrix.
Obviously, the complete determination of
the $\mathbf{M}$ amounts to solving the
Schr\"{o}dinger equation and, in consequence,
it is not, in general, a simple exercise.
Nevertheless, some properties of the transfer
matrix are universal~\cite{Grif01}. First,
we note that time-reversal invariance implies
[because $V(x)$ is real] that $\psi^\ast(x)$
is also a solution. Since this symmetry
interchanges incoming and outgoing waves
this means that
\begin{equation}
\label{Mast}
\left (
\begin{array}{c}
A_-^\ast \\
A_+^\ast
\end{array}
\right )
=
\mathbf{M}
\left (
\begin{array}{c}
B_-^\ast \\
B_+^\ast
\end{array}
\right ) .
\end{equation}
Comparing with Eq.~(\ref{M}) leads to the
conclusion that the matrix $\mathbf{M}$
must be of the form
\begin{equation}
\label{Mab}
\mathbf{M} =
\left (
\begin{array}{cc}
\alpha & \beta \\
\beta^\ast & \alpha^\ast
\end{array}
\right ) .
\end{equation}

Next we assess the implications of the
conservation of probability. Since the
probability current is
\begin{equation}
J = - i \frac{\hbar}{2 m}
\left ( \psi^\ast \frac{d\psi}{dx} -
\psi \frac{d\psi^\ast}{dx} \right ) ,
\end{equation}
the continuity equation entails
\begin{equation}
\label{fluxcon}
|A_+|^2 - |A_-|^2 = |B_+|^2 - |B_-|^2 ,
\end{equation}
which is tantamount to
\begin{equation}
\label{det1}
\det \mathbf{M} = 1 .
\end{equation}
The set of $2 \times 2$ complex matrices of
the form (\ref{Mab}) satisfying the constraint
(\ref{det1}) constitute a group called
SU(1,~1)~\cite{Wybo74}.

If we take an incident wave from the left
($B_-=0$) and fix $A_+=1$, then
\begin{equation}
\left (
\begin{array}{c}
1 \\
r
\end{array}
\right )
=
\mathbf{M}
\left (
\begin{array}{c}
t \\
0
\end{array}
\right ) ,
\end{equation}
where the complex numbers $r$ and $t$ are
the corresponding reflection and transmission
amplitudes. This determines the first column
of $\mathbf{M}$. Denoting the corresponding
amplitudes for waves incident from the right as
$r^\prime$ and $t^\prime$ and repeating
the procedure, one easily finds that
time-reversal invariance imposes
\begin{eqnarray}
& t^\prime = t , & \nonumber \\
& & \\
& \displaystyle
\frac{r^\prime}{t^\prime} = -
\frac{r^\ast}{t^\ast} , & \nonumber
\end{eqnarray}
while conservation of the flux determines
\begin{equation}
|r|^2 + |t|^2 = 1 . \nonumber \\
\end{equation}

The final form of our transfer matrix
is then
\begin{equation}
\label{TranM}
\mathbf{M} =
\left(
\begin{array}{cc}
1/t & r^{\ast}/t^{\ast} \\
r/t & 1/t^{\ast}
\end{array}
\right) .
\end{equation}
In the particular case of a symmetric
potential it is obvious that $r = r^\prime$
and therefore the matrix element $\beta$ is
an imaginary number.

For later use, we will now bring up the
paradigmatic example of a rectangular
potential barrier of width $L$ and height
$V_0$. Since the calculations can be
easily carried out, we skip the details
and merely quote the results for $r$ and
$t$:
\begin{eqnarray}
\label{Barhyper}
r & = & e^{+ i k L}
\left [
\frac{(k^2 + \kappa^2) \sinh(\kappa L)}
{(k^2- \kappa^2) \sinh (\kappa L)
+ 2 i k \kappa \cosh (\kappa L)}
\right ] , \nonumber \\
& & \\
t & = &
e^{- i k L}
\left [
\frac{2 i k \kappa}
{(k^2 - \kappa^2) \sinh (\kappa L)
+ 2 i k \kappa \cosh (\kappa L)} \right ] ,
\nonumber
\end{eqnarray}
where $k^2 = 2 m E /\hbar^2$ and
$\kappa^2 = 2m (E - V_0)/\hbar^2$. These
coefficients correspond to the case
$E < V_0$. When $E > V_0$ the above
expressions remain valid with the formal
substitution $\kappa \rightarrow i
\bar{\kappa}$. Finally, when $E = V_0$,
a limiting procedure gives
\begin{eqnarray}
\label{Barpar}
r & = & e^{+ i kL}
\left [
\frac{1}{1 + 2i / (kL)}
\right ]\, ,
\nonumber \\
& & \\
t & = &
e^{- i k L}
\left [
\frac{1}
{1 + k L/(2i)} \right ] \, .
\nonumber
\end{eqnarray}

Be aware that the transfer matrix depends
on the choice of basis vectors. For example,
instead of specifying the amplitudes of the
right and left-moving waves, we could write
a linear relation between the values of the
wave function and its derivative at two
different points~\cite{Spru93}:
\begin{equation}
\left (
\begin{array}{c}
\psi(a) \\
\psi^\prime (a)
\end{array}
\right )
=
\bm{\mathcal{M}}
\left (
\begin{array}{c}
\psi(b) \\
\psi^\prime (b)
\end{array}
\right ) .
\end{equation}
These two basis vectors are related by
\begin{equation}
\left (
\begin{array}{c}
\psi(a) \\
\psi^\prime (a)
\end{array}
\right )
=
\bm{\mathcal{U}}
\left (
\begin{array}{c}
A_+ \\
A_-
\end{array}
\right ) ,
\end{equation}
where
\begin{equation}
\bm{\mathcal{U}} =
\left(
\begin{array}{cc}
1 & 1 \\
i k & - i k
\end{array}
\right) ,
\end{equation}
and analogously at the point $b$. Correspondingly,
the matrix $\mathbf{M}$ in this representation is
\begin{equation}
\bm{\mathcal{M}}
= \bm{\mathcal{U}}
\mathbf{M}
\bm{\mathcal{U}}^{-1}
=
\left(
\begin{array}{cc}
\mathfrak{a} & \mathfrak{b} \\
\mathfrak{c} & \mathfrak{d}
\end{array}
\right ) ,
\end{equation}
where
\begin{eqnarray}
& \mathfrak{a} = \mathrm{Re}\, \alpha +
\mathrm{Re} \, \beta ,
\qquad
\displaystyle
\mathfrak{b} = \frac{1}{k} (\mathrm{Im} \, \alpha -
\mathrm{Im} \, \beta ) , & \nonumber \\
& & \\
& \mathfrak{c} = - k (\mathrm{Im} \, \alpha +
\mathrm{Im} \, \beta ) \, ,
\qquad
\mathfrak{d} = \mathrm{Re} \, \alpha -
\mathrm{Re} \, \beta \, . &
\nonumber
\end{eqnarray}
Since the trace and the determinant are preserved
by this matrix conjugation, we have that
$\det \bm{\mathcal{M}}= +1$. In consequence,
in this representation transfer matrices
belong to the group SL(2, $\mathbb{R}$) of
unimodular $2 \times 2$ matrices with real
elements.

Transfer matrices are very convenient mathematical
objects. Suppose we know how the wave functions
``propagate" from point $b$ to point $a$, with
a transfer matrix we symbolically write as
$\mathbf{M}(a,b)$, and also from $c$ to $b$,
with $\mathbf{M}(b, c)$. The essential point
is that propagation from $c$ to $a$ is then
described by the product of transfer matrices:
\begin{equation}
\mathbf{M} (a, c) = \mathbf{M} (a, b) \,
\mathbf{M} (b, c) .
\end{equation}
The multiplicative property is rather useful:
we can connect simple scatterers as building
blocks to create an intricate potential
landscape and determine its transferº matrix
by simple multiplication. The usual scattering
matrix does not have this important property
because the incoming amplitudes for the overall
system cannot be obtained in terms of the
incoming amplitudes for every subsystem.

\section{Understanding scattering amplitudes
in the unit disk}

We observe that because of the flux conservation
in Eq.~(\ref{fluxcon}), the complex quotients
\begin{equation}
\label{defz}
z_a = \frac{A_-}{A_+} \, ,
\qquad
z_b = \frac{B_-}{B_+} \, ,
\end{equation}
contain the essential information about the
wave function and omit a global phase factor.
The action of a transfer matrix can be then
seen as a mapping from the value on $z_b$
to the value on $z_a$ according to
\begin{equation}
\label{accion}
z_a = \Phi [\mathsf{M} , z_b] =
\frac{\beta^\ast +\alpha^\ast z_b}
{\alpha + \beta z_b} \ ,
\end{equation}
which can be appropriately called the
scattering transfer function~\cite{Monz02}.
The action (\ref{accion}) is known as a
bilinear or M\"obius mapping and is a
conformal mapping of the entire plane
onto itself and which maps circles into circles.

Properties of this mapping are part of any
course in complex variables and have been
discussed, in the context of a relativisticlike
presentation of multilayer optics, in
Refs.~\cite{Yont02}~and~\cite{Monz02}. One can
check that points in the unit disk are mapped
onto points in the unit disk, while the unit
circle maps into itself. The external
region remains also invariant. For the usual
scattering solution with incident waves from
the left ($B_- = 0$), we have $z_b = 0$ and
$z_a = r$. Conversely, when $z_a = 0$, we have
the necessary and sufficient condition for a
transparent potential. Note that the unit
circle represents the action of a system
with $| r | = 1$, that is, of a perfect
``mirror".

To classify the scatterer action it proves
convenient to work out the fixed points of
the mapping~\cite{Ande99}; that is, the wave
configurations such that $z_a = z_b \equiv z_f$
in Eq.~(\ref{accion}):
\begin{equation}
z_f = \Phi [\mathbf{M}, z_f] ,
\end{equation}
whose solutions are
\begin{equation}
z_{f\pm} = \frac{1}{2 \beta}
\left \{ -2 i \ \mathrm{Im}(\alpha) \pm
\sqrt{[\Tr ( \mathbf{M} )]^2 -4}
\right \} .
\end{equation}
When $ [\Tr ( \mathbf{M} )] ^2 < 4$
the action is said elliptic and has only
one fixed point inside the unit disk.
Since in the Euclidean geometry a rotation
is characterized for having only one invariant
point, this action can be appropriately
called a hyperbolic rotation.

When $ [ \Tr ( \mathbf{M})]^2 > 4$
the action is said hyperbolic and has two
fixed points, both on the unit circle.
The geodesic line joining these two fixed points
remains invariant and thus, by analogy with the
Euclidean case, this action will be called a
hyperbolic translation.

Finally, when $ [ \Tr ( \mathbf{M}) ]^2 = 4$
the system action is parabolic and has only
one (double) fixed point on the unit circle.

It is worth mentioning that, for the example
of the rectangular barrier discussed previously,
one can use Eqs.~(\ref{Barhyper}) and (\ref{Barpar})
to check that its action becomes elliptic,
hyperbolic, or parabolic according to $E$
is greater than, lesser than, or equal to
$V_0$, respectively.

To proceed further let us note that by taking
the conjugate of $\mathbf{M}$ with any matrix
$\mathbf{C} \in $ SU(1, 1), i. e.,
\begin{equation}
\label{conjC}
\mathbf{M}_{\mathrm{C}} = \mathbf{C} \ \mathbf{M} \
\mathbf{C}^{-1} ,
\end{equation}
we obtain another matrix of the same type,
since $\Tr ( \mathbf{M} ) = \Tr (
\mathbf{M}_{\mathrm{C}} )$. Conversely, if two
systems have the same trace, one can always find
a matrix $\mathbf{C}$ satisfying Eq.~(\ref{conjC}).

The fixed points of $\mathbf{M}_{\mathrm{C}}$ are
then the image by $\mathbf{C}$ of the fixed points
of $\mathbf{M}$. In consequence, given any
transfer matrix $\mathbf{M}$ we can always
reduce it to one of the following canonical
forms~\cite{Sanc01}:
\begin{eqnarray}
\label{Iwasa}
\mathbf{K}_{\mathrm{C}} (\theta ) & = &
\left (
\begin{array}{cc}
\exp (i\theta/2) & 0 \\
0 & \exp (-i\theta/2)
\end{array}
\right ) \, ,
\nonumber \\
\mathbf{A}_{\mathrm{C}} (\xi) & = &
\left (
\begin{array}{cc}
\cosh (\xi/2) & i\, \sinh(\xi/2) \\
-i\, \sinh(\xi/2) & \cosh (\xi/2)
\end{array}
\right ) \ , \\
\mathbf{N}_{\mathrm{C}} ( \nu ) & = &
\left (
\begin{array}{cc}
1 - i \nu/2 & \nu/2 \\
\nu/2 & 1+ i \nu/2
\end{array}
\right ) \, ,
\nonumber
\end{eqnarray}
which have as fixed points the origin
(elliptic), $+i$ and $-i$ (hyperbolic)
and $+i$ (parabolic), respectively.

For the canonical forms (\ref{Iwasa}) the
corresponding bilinear transformations are
\begin{eqnarray}
\label{orb}
z^\prime & = &
\Phi [\mathbf{K}_{\mathrm{C}} (\theta), z ]
= z \exp (-i\theta) \ , \nonumber \\
& & \nonumber \\
z^\prime & = &
\Phi [\mathbf{A}_{\mathrm{C}} (\xi), z ]
= \frac{z - i \tanh(\xi/2)}
{1 + i z \tanh(\xi/2)} \, , \\
& & \nonumber \\
z^\prime & = &
\Phi [\mathbf{N}_{\mathrm{C}} (\nu), z ]
= \frac{z +(1+i z) \nu /2}
{1 + (z - i) \nu/2} \, . \nonumber
\end{eqnarray}
In words, given a generic point $z$ in the
unit disk, and varying the parameters $\theta$,
$\xi$, or $\nu$ in Eq.~(\ref{orb}), the transformed
points $z^\prime$ describe a characteristic
curve that we shall call the orbit associated
to $z$ by the transformation. In Fig.~1.a we
have plotted some orbits for different values
of $z$ for each one of these canonical forms.
For matrices $\mathbf{K}_{\mathrm{C}}
(\theta)$ the orbits are circumferences
centered at the origin. For the matrices
$\mathbf{A}_{\mathrm{C}} (\xi)$, they
are arcs of circumference going from the
point $ +i$ to the point $-i$ through $z$.
Finally, for the matrices $\mathbf{N}_{\mathrm{C}}
(\nu)$ the orbits are circumferences passing
through the points $i$, $z$, and $-z^\ast$.
Note that this is in full agreement with
the geometrical meaning of these transformations.
In Fig.~1.b we have plotted the corresponding
orbits for arbitrary fixed points, obtained
by conjugation of the previous ones. The
explicit construction of the family of matrices
$\mathbf{C}$ is not difficult: it suffices
to impose that $\mathbf{C}$ transforms
the fixed points of $\mathbf{M} $ into
the ones of $\mathbf{K}_{\mathrm{C}} (\theta)$,
$\mathbf{A}_{\mathrm{C}} (\xi)$, or
$\mathbf{N}_{\mathrm{C}} (\nu)$, respectively.

\begin{figure}
\centering
\resizebox{0.80\columnwidth}{!}{\includegraphics{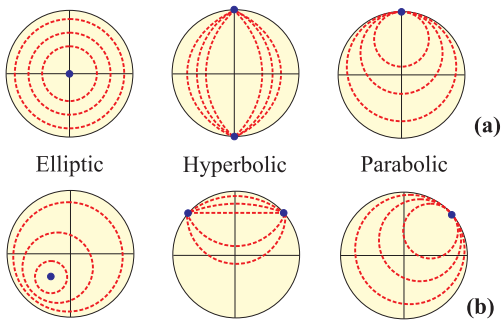}}
\caption{Plot of typical orbits in the unit disk:
(a) canonical transfer matrices as given in Eq.~(\ref{Iwasa})
and (b) arbitrary transfer matrices obtained by
conjugation as in Eq.~(\ref{conjC}) of the previous
ones.}
\end{figure}

\section{Application: geometrical representation
of finite periodic systems}

As an important application of the previous
formalism, let us suppose that we repeat $N$
times our system represented by $\mathbf{M}$.
This is called a finite periodic system
and, given its relevance, has been
extensively discussed in the
literature~\cite{Vezz86,Kalo91,Griff92,Rozm94,Chup94,Livi94,Erdo97,Barr99}.
Obviously, the overall transfer matrix
is now $\mathbf{M}^N$, so all the
algebraic task reduces to the obtention
of a closed expression for the $N$th power
of the matrix $\mathbf{M}$. Although there
are several elegant ways of computing this,
we shall instead apply our geometrical picture.
To this end we represent the transformed
state by the $N$-period structure by the point
\begin{equation}
\label{iterat1}
z_N = \Phi[\mathbf{M}, z_{N-1}] =
\Phi[\mathbf{M}^N, z_0] ,
\end{equation}
where $z_0$ denotes here the initial point.

Henceforth, we shall take $z_0 = 0$, which
is not a serious restriction as it corresponds
to the case in which no wave incides from the
right. Note also that all the points $z_N$
lie in the orbit associated to the initial
point $z_0$ by the basic period, which is
determined by its fixed points: the character
of these fixed points determine thus the
behavior of the periodic structure.

To illustrate how this geometrical approach
works in practice, in Fig.~3 we have plotted
the sequence of successive iterates obtained
numerically for different kind of transfer
matrices according to our previous classification.

In the elliptic case, the points $z_N$
revolve in the orbit centered at the
fixed point and the system never
reaches the unit circle. On the contrary,
for the hyperbolic and parabolic cases
the iterates converge to one of the fixed
points on the unit circle, although with
different laws. Since the unit circle
represents a perfect ``mirror", this means
that strong reflection occurs and we are in
a forbidden band. In other words, in this
geometrical picture the route to a forbidden
band can be understood as the convergence of
the point representing the action of the system
to the unit circle.

Obviously, this is in perfect agreement with
the standard treatment, which gets these band
gaps from an eigenvalue equation for the Bloch
factor in an infinite periodic structure: since
the Bloch phase is $ | \Tr ( \mathbf{M} ) |$,
and strong reflection occurs when this trace
exceeds 2 in magnitude (the band edge is located
precisely when the trace equals 2)~\cite{Barr03}.

\begin{figure}
\centering
\resizebox{0.80\columnwidth}{!}{\includegraphics{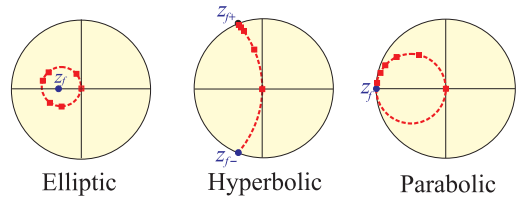}}
\caption{Plot of the successive iterates ($N = 1, \ldots, 5$)
for an elliptic, hyperbolic, and parabolic action starting
from the origin as the initial point. Only hyperbolic and
parabolic actions tend to the unit circle.}
\end{figure}

Let us focus then on the hyperbolic case,
which, in this approach, corresponds to
a translation in the unit disk. We can
explicitly compute the $N$th iterate for
the canonical form $\mathbf{A}_{\mathrm{C}}$,
since $\mathbf{A}_{\mathrm{C}}(\xi_1)
\mathbf{A}_{\mathrm{C}} (\xi_2) =
\mathbf{A}_{\mathrm{C}} (\xi_1 + \xi_2)$
[this property holds true for all the canonical
forms in Eq.~(\ref{Iwasa})]. Finally, it suffices
to conjugate as in (\ref{conjC}) to obtain,
after some calculations, that
\begin{equation}
z_N = \frac{1- \chi^N }{1 -
\chi^N  (z_{f +}/z_{f -})}z_{f +} ,
\end{equation}
where
\begin{equation}
\chi = \frac{\alpha + \beta z_{f -}}
{\alpha + \beta z_{f +}}
\end{equation}
is a complex number satisfying $| \chi | < 1$.
Here $z_{f \pm}$ are the fixed points of
the matrix. Note that, because $z_0 = 0$,
this initial point is transformed by the single
period into the point $r$. Therefore, $z_N$
represents the reflection amplitude of
the overall periodic structure, which is
obviously different from $r^N$. One gets
\begin{equation}
|z_N|^2 =
\frac{ | \beta|^2}{|\beta|^2 +
[\sinh ( \xi) / \sinh (N  \xi ) ]^2 },
\end{equation}
where we have denoted
\begin{equation}
\mathrm{Re} \, \alpha = \frac{1}{2}
\Tr (\mathbf{M}) = \cosh( \xi) .
\end{equation}
Note that $|z_N|^2$ approaches the unit
circle exponentially with $N$, as one
could expect from a band stop.

Analogously, for the parabolic case we have
\begin{equation}
z_N = \frac{N \beta z_f^2}{ N \beta z_f - 1} ,
\end{equation}
where $z_f$ is the (double) fixed point. One
then obtains
\begin{equation}
|z_N|^2 =
\frac{ | \beta|^2}{| \beta|^2 + ( 1/ N ) ^2 } ,
\end{equation}
that goes to unity with a typical behavior
$ O(N^{-2})$. This is universal in the physics
of reflection, as put forward in a different
framework by Lekner~\cite{Lekn87} and Yeh~\cite{Yeh88}.

\section{Transfer-matrix composition as a
hyperbolic-turn sum}

To clarify the geometrical picture of the
composition of two scattering systems we
briefly recall that the (hyperbolic) metric
in the unit disk is defined by~\cite{Ande99}
\begin{equation}
\label{metricH}
ds^2 = \frac{dz^2}{(1 - |z|^2)^2} .
\end{equation}
With this metric, it is a simple exercise to
work out that the geodesic (path of minimum
distance) between two points is the Euclidean
arc of the circle through those points
and orthogonal to the unit circle. The
diameters are also geodesic lines.

The hyperbolic distance between two
points $z_b$ and $z_a$ can be computed
from (\ref{metricH}) and is
\begin{equation}
d_{\mathbb{H}} (z_b, z_a) = \ln \left (
\frac{|z_b^\ast z_a - 1 | + |z_b - z_a|}
{|z_b^\ast z_a - 1 |- |z_b - z_a |} \right ) \, .
\end{equation}
Note that the visual import of the disk with
this metric is that a pair of points with
a given distance between them will appear
to be closer and closer together as their
location approaches the boundary circle. Or,
equivalently, a pair of points near the boundary
of the disk are actually farther apart (via
the metric) than a pair near the center of
the disk which appear to be the same distance
apart.

When these points $z_b$ and $z_a$ are related by
$\mathbf{M}$, the distance between them can
be compactly expressed as
\begin{equation}
\zeta = 2 \ln  \left (
\frac{1}{2} \{ \Tr ( \mathbf{M}) +
\sqrt{[\Tr ( \mathbf{M} ) ]^2 - 4} \}
\right ) \, .
\end{equation}

Let us focus on the case of $ [ \Tr (
\mathbf{M} )]^2 > 4$. This is not a
serious restriction, since it is known
that any matrix of SU(1,~1) can be written
(in many ways) as the product of two hyperbolic
translations~\cite{Simo89b}. The axis of the
hyperbolic translation is the geodesic line
joining the two fixed points.

\begin{figure}
\centering
\resizebox{0.60\columnwidth}{!}{\includegraphics{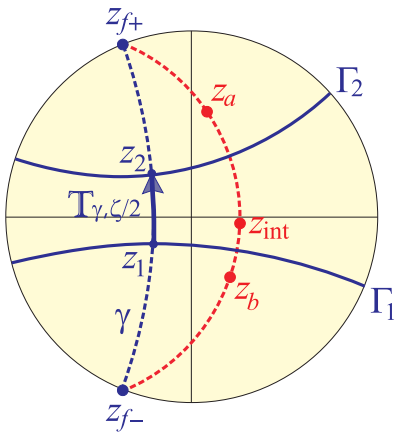}}
\caption{Representation of the sliding turn
$\mathbb{T}_{\gamma, \zeta/2}$ in terms of two
reflections in two lines $\Gamma_1$ and $\Gamma_2$
orthogonal to the axis of the translation $\gamma$,
which has two fixed points $z_{f+}$ and $z_{f-}$. The
transformation of a typical off axis point $z_b$ is also shown.}
\end{figure}

In Euclidean geometry, a translation of
magnitude $\zeta$ along a line $\gamma$
can be seen as the product of two reflections
in any two straight lines orthogonal to
$\gamma$, separated a distance $\zeta/2$.
This idea can be translated much in the
same way to the unit disk, once the
concepts of line and distance are understood
in the hyperbolic sense. In consequence,
any pair of points $z_1$ and $z_2$ on
the axis of the translation $\gamma$ at a
distance $\zeta/2$ can be chosen as intersections
of $\Gamma_1$ and $\Gamma_2$ (orthogonal lines
to $\gamma$) with $\gamma$. It is then
natural to associate to the translation
an oriented segment of length $\zeta/2$ on
$\gamma$, but otherwise free to slide on
$\gamma$ (see Fig.~4). This is analogous
to Hamilton's turns, and will be called
a hyperbolic turn $\mathbb{T}_{\gamma, \zeta/2}$.

Note that using this construction, an off-axis
point such as $z_b$ will be mapped by these
two reflections (through an intermediate
point $z_{\mathrm{int}}$) to another point
$z_a$ along a curve equidistant to the axis.
These other curves, unlike the axis of translation,
are not hyperbolic lines. The essential point
is that once the turn is known, the transformation
of every point in the unit disk is automatically
established.

Alternatively~\cite{Barr04}, we can formulate
the concept of turn as the ``square root"
of a transfer matrix: if $\mathbf{M}$ is
a hyperbolic translation with $ \Tr (
\mathbf{M} ) $ positive (equivalently,
$\mathrm{Re} (\alpha ) > 1$), then one can ensure
that its square root exists and reads as
\begin{equation}
\sqrt{\mathbf{M}} =
\frac{1}{\sqrt{2 [ \mathrm{Re} (\alpha )
+ 1 ]}}
\left [
\begin{array}{cc}
\alpha + 1 & \beta \\
\beta^\ast & \alpha^\ast + 1
\end{array}
\right ] .
\end{equation}
We can easily check that this matrix has
the same fixed points as $ \mathbf{M}$,
but the translated distance is just
half the induced by $ \mathbf{M}$;
that is
\begin{equation}
\zeta_{\mathbf{M}} =
2 \zeta_{\sqrt{\mathbf{M}}} .
\end{equation}
This suggests that the matrix $ \sqrt{\mathbf{M}}$
can be appropriately associated to the turn
$\mathbb{T}_{\gamma, \zeta/2}$ that represents
the translation induced by $\mathbf{M}$.

One may be tempted to extend the Euclidean
composition of concurrent vectors to the
problem of hyperbolic turns. Indeed, this
can be done quite straightforwardly~\cite{Juar82}.
Let us consider the case of the composition
of two of these systems represented by
matrices $\mathbf{M}_1$ and $\mathbf{M}_2$
with scattering amplitudes $(r_1, t_1)$ and
$(r_2, t_2)$, in agreement with Eq.~(\ref{TranM}).
The action of the compound system can be expressed as
\begin{equation}
\mathbf{M}_{12} = \mathbf{M}_1
\mathbf{M}_2 \, ,
\end{equation}
and the reflection and transmission amplitudes
associated to $\mathbf{M}_{12}$ are
\begin{eqnarray}
\label{12}
r_{12} & = & \frac{r_1 + r_2
\exp (i 2 \varphi_1)}
{1 + r_1^\ast r_2 \exp (i 2 \varphi_1)} ,
\nonumber \\
& & \\
t_{12} & = & \frac{t_1 t_2}
{1 + r_1^\ast r_2 \exp (i 2 \varphi_1)} ,
\nonumber
\end{eqnarray}
where $t_1 = |t_1 | \exp(i \varphi_1)$.

Let $\zeta_1$ and $\zeta_2$ be the corresponding
translated distances along intersecting axes
$\gamma_1$ and $\gamma_2$, respectively. Take
now the associated turns $\mathbb{T}_{\gamma_1,
\zeta_1/2}$ and $\mathbb{T}_{\gamma_2, \zeta_2 /2}$
and slide them along $\gamma_1$ and $\gamma_2$ until
they are ``head to tail". Afterwards, the turn
determined by the free tail and head is the turn
associated to the resultant, which represents
thus a translation of parameter $\zeta_{12}$ along the
line $\gamma_{12}$.

\begin{figure}
\centering
\resizebox{0.60\columnwidth}{!}{\includegraphics{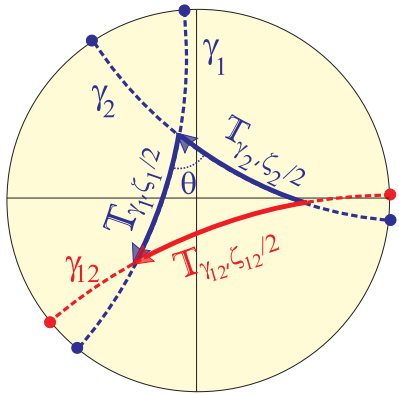}}
\caption{Composition of two scattering systems represented
by the hyperbolic turns $\mathbb{T}_{\gamma_1, \zeta_1/2}$
and $\mathbb{T}_{\gamma_2, \zeta_2/2}$. The action of the
overall system is obtained  by using a parallelogramlike law.}
\end{figure}

This construction is shown in Fig.~5, where
the pertinent parameters are $(r_1 = -0.9521 - i 0.0882, t_1 =
0.2532 - i 0.1468)$ and $(r_2 = -0.3307- i 0.52903, t_2 = 0.6284 -
i 0.4647)$. The application of (\ref{12}) gives
$(r_{12} = 0.3736 + i 0.2014, t_{12} = 0.8971 - i 0.1228)$. The
noncommutative character is evident, and can
also be inferred from the obvious fact that
$\mathbf{M}_{12} \neq \mathbf{M}_{21}$.

In Euclidean geometry, the resultant of this parallelogram
law can be quantitatively determined by a direct application
of the cosine theorem. For any hyperbolic triangle with sides
of lengths $\zeta_1$ and $\zeta_2$ that make an angle $\theta$,
we take the expression from any standard book on hyperbolic
geometry~\cite{Ande99}
\begin{equation}
\label{hcos}
\cosh \zeta_{12} = \cosh \zeta_1
\cosh \zeta_2 + \sinh \zeta_1
\sinh \zeta_2 \cos \theta ,
\end{equation}
where $\theta$ is the angle between both sides.

\section{Concluding remarks}

In summary, what we hope to have accomplished is
to present in a clear way the advantages of using
the transfer matrix to study one-dimensional
scattering. In spite of the slight ``cross-talk"
between different fields, the transfer matrix is
a powerful tool that relies only on linearity
of a nonabsorbing system with two input and
two output channels. For this reason, it
is becoming more and more important in a
variety of applications.

We have interpreted the action of a transfer
matrix on a wave function as a conformal mapping
on the unit disk and we have characterized
the basic geometrical actions in terms of its
trace. By generalizing to the unit disk Hamilton's
idea of turns, we have provided a remarkably
vivid pictorial description of the scattering
action, with a composition law that parallels
the corresponding one for sliding vectors in
Euclidean geometry.

To conclude, we expect that the geometrical
scenario presented here could provide an
appropriate tool for analyzing the
performance of one-dimensional potentials
in an elegant and concise way.

\section*{Acknowledgments}

Our efforts towards understanding the problems
posed in this paper were fueled in part, and were
made much more interesting, by the interaction
with Alberto Galindo. For this and many other
reasons, it is a pleasure to dedicate this
paper to Alberto on the occasion of his
seventieth birthday.

We would like to thank Gunnar Bjork, Hubert de
Guise, Andrei Klimov, Jos\'e Mar\'{\i}a Montesinos
and Ma\-ria\-no Santander for valuable discussions on
different aspects of scattering in one dimension.

\end{document}